\documentclass[12pt,letterpaper]{iopart}
\usepackage{amsmath,amssymb}
\usepackage{color}
\usepackage{cite}

\let\Tr\undefined
\usepackage[fonts]{faust}

\usepackage[linkcolor=blue, colorlinks=true, citecolor=red]{hyperref}

\newcommand\M{\mathcal{M}}
\newcommand\ED{\mathrm{D}}

\begin{document}
\title{A note on axial symmetries}
\author{Christopher Beetle and Shawn Wilder}
\address{Department of Physics, Florida Atlantic University, 777 Glades Road, Boca Raton, Florida 33431}

\begin{abstract}
This note describes a local scheme to characterize and normalize an axial Killing field on a general Riemannian geometry.  No global assumptions are necessary, such as that the orbits of the Killing field all have period $2 \pi$.  Rather, any Killing field that vanishes at at least one point necessarily has the expected global properties.
\end{abstract}

\section{Introduction}
 
Axial Killing fields play an important role in classical general relativity when defining the angular momentum of a black hole or a similar compact, gravitating body.  The simplest example is the Komar angular momentum, which applies to regions of spacetime that admit a global axial Killing field $\varphi^a$.  The formula is 
\begin{equation}\label{Komar}
	J_{\text{Komar}} 
		:= \frac{1}{16 \pi} \oint_S \grad\!_a\, \varphi_b\, \ed S^{ab}
		= \frac{1}{8 \pi} \oint_S K_{ab}\, s^a\, \varphi^b\, \epsilon, 
\end{equation}
where $S$ is a spacelike 2-sphere, $\ed S^{ab}$ is the area bivector normal to  $S$, $K_{ab}$ is the extrinsic curvature of a Cauchy surface $\Sigma$ containing $S$, $s^a$ is the spacelike normal to $S$ within $\Sigma$, and $\epsilon$ is the intrinsic area element on $S$.  Similar integrals --- such as the quasi-local formulae due to Brown and York \cite{BY:QLoc} and for dynamical horizons \cite{AK:Rev} --- apply more generally, when only the \textit{intrinsic, two-dimensional} metric on $S$ is symmetric.  But the definition of each does rely on the existence of such an intrinsic symmetry.

In order to justify an integral like (\ref{Komar}) as a physical angular momentum, $\varphi^a$ should have some standard, global characteristics of an axial Killing field.  In particular, for a general Riemannian geometry $(\M, g_{ab})$, these characteristics include that 
\begin{enumerate}\renewcommand\theenumi{\alph{enumi}}
\item $\varphi^a$ vanishes at at least one fixed point $p_0 \in \M$, 
\item every orbit of the flow $\Phi(t)$ generated by $\varphi^a$ is either a fixed point or a circle, and 
\item all of the circular orbits close only at integer multiples of a common period $t = t_0$.
\end{enumerate}
One may then scale $\varphi^a$ by a constant as needed so that $t_0 = 2 \pi$, the standard period for a group of rigid rotations.  The angular momenta mentioned above take unique values only when one restricts to this preferred normalization.  If $\M \sim S^2$ happens to be a (topological) 2-sphere, then condition (a) is usually strengthened to 
\begin{enumerate}
\item[(A)] $\varphi^a$ vanishes at exactly two fixed points $p_0$ and $p_1$ in $\M \sim S^2$.
\end{enumerate}
These conditions distill the intuitive, geometric features of axial symmetries in standard geometries (\textit{i.e.}, Euclidean, Gaussian, or Lobachevskian).

The purpose of this note is to show that the conditions above are redundant.  The \textit{global} conditions (b) and (c) --- as well as condition (A) when $\M \sim S^2$ --- derive from condition (a) and one other \textit{local} condition at $p_0$, which is that 
\begin{equation}\label{Fcon}
	\bigl( \Tr F^2 \bigr) F^3 = \bigl( \tr F^4 \bigr) F,
	\qquad\text{where}\qquad
	F^b{}_a := \grad\!_a\, \varphi^b \bigr|_{p_0}.
\end{equation}
That is, any Killing field $\varphi^a$ with at least one fixed point $p_0$, whose derivative satisfies the local condition (\ref{Fcon}), is necessarily an axial Killing field in the global sense described above.  Furthermore, (\ref{Fcon}) \textit{always} holds if $\M$ is 2- or 3-dimensional, so in those cases any Killing field with a fixed point is an axial Killing field.  Finally, every vector field on a 2-sphere must vanish at at least one point, so every Killing field on a 2-sphere is axial.  Condition (A) then follows by a topological argument.

One can also show that conditions (A), (b), and (c) hold whenever $\M \sim S^2$ is a 2-sphere \cite[see \S 3]{ABL:Mech} using the classical uniformization theorem: every metric $g_{ab}$ on a 2-sphere is conformally related to a round metric $\mathring g_{ab}$.  Any Killing field of $g_{ab}$ must also be a Killing field of $\mathring g_{ab}$, and therefore must satisfy conditions (A), (b), and (c).  The approach we take here is more direct, and actually lays the foundation \cite{CLT:Unif} for a proof of the uniformization theorem via Ricci flow \cite{C:Sph}.  Here we extend this direct approach to arbitrary Riemannian geometries and highlight its utility in physical applications.

\section{From Global to Local}

The global flow $\Phi(t)$ generated by a Killing field $\varphi^a$ on a Riemannian manifold $(\M, g_{ab})$ is closely related to the corresponding inifnitesimal flow $\Phi'(t)$ induced on the tangent space $T_{p_0} \M$ at any point $p_0$ where $\varphi^a$ vanishes.  The key point is that the latter form a subgroup of the symmetry group $SO(n)$ for the Euclidean geometry $\bigl( T_{p_0} \M, g_{ab}(p_0) \bigr)$.  

The natural flow $\Phi'(t)$ in the tangent space at $p_0$ consists of the differentials $\ED_{p_0} \Phi(t)$ of the diffeomorphisms $\Phi(t)$ in the global flow on $\M$.  The differential of a diffeomorphism $\Psi : \M \to \M$ is just the natural push-forward map 
\begin{equation}
	\ED_p \Psi := \Psi_* \bigr|_p : T_p \M \to T_{\Psi(p)} \M
\end{equation}
on vectors.  If $\Psi(p) = p$ is fixed, then $D_p \Psi$ becomes a linear map from the tangent space $T_p \M$ to itself.  Furthermore, if $\Psi$ is a global isometry of the Riemannian geometry $(\M, g_{ab})$ that fixes $p \in \M$, then $D_p \Psi$ is a linear isometry of the Euclidean geometry $\bigl( T_p \M, g_{ab}(p) \bigr)$.  It follows that 
\begin{equation}
	\Phi'(t) := \ED_{p_0} \Phi(t) := \Phi_*(t) \bigr|_{p_0} : T_{p_0} \M \to T_{p_0} \M
\end{equation}
is a one-parameter group of linear isometries of a Eucliean vector space.  They must be rotational isometries, not translational, because they fix the origin of $T_{p_0} \M$.  Thus, each orbit of $\Phi'(t)$ must either be a fixed point or a circle.  This is the tangent-space analogue (b$'$) of condition (b) above.

Next, recall that the Lie derivative along $\varphi^a$ of a vector field $v^a$ is defined in terms of the push-forward under the flow $\Phi(t)$ generated by $\varphi^a$: 
\begin{equation}
	\Lie_\varphi\, v \bigr|_p := \pdby{t}\, \Bigl[ \Phi_*(-t)(v) \bigr|_p \Bigr]_{t = 0}.
\end{equation}
When $\varphi^a$ vanishes at $p = p_0$, this determines the generator, in the ordinary sense of matrices, of the one-parameter group of linear isometries $\Phi'(t)$ via 
\begin{equation}
	\pdby{t}\, \Bigl[ \Phi'(t)^b{}_a\, v^a \Bigr]_{t = 0} 
		= - \Lie_\varphi\, v^b \bigr|_{p_0}
		= v^a\, \grad\!_a\, \varphi^b \bigr|_{p_0}.
\end{equation}
Thus, the tensor $F^b{}_a$ from (\ref{Fcon}) generates the rotations $\Phi'(t)$ in $T_{p_0} \M$ in the sense that 
\begin{equation}\label{expF}
	\Phi'(t) = \ee^{t F}.
\end{equation}
One can always find an orthonormal basis for $\bigl( T_{p_0} \M, g_{ab}(p_0) \bigr)$ that puts an anti-symmetric tensor like $F^b{}_a$ in canonical, block-diagonal form with one or more $2 \times 2$ anti-symmetric blocks along the diagonal, possibly followed by one or more zeroes.  There can be at most one non-zero block if $\M$ is 2- or 3-dimensional, but in higher dimensions there may be several.  In principle, each block could have a different pair of values $\pm f_i$ for its off-diagonal elements.  But the period of $\ee^{t F}$ within the 2-dimensional subspace of $T_{p_0} \M$ corresponding to one of these blocks is $2 \pi / f_i$.  These periods are equal if (\ref{Fcon}) holds because, working in the preferred basis, we have 
\begin{equation}\label{fiVals}
	\Bigl( \sum\nolimits_j - 2 f_j^2 \Bigr) \cdot - f_i^3 = \Bigl( \sum\nolimits_j 2 f_j^4 \Bigr)\, f_i
	\qquad\Leftrightarrow\qquad
	f_i^2 = \frac{\sum_j f_j^4}{\sum_j f_j^2} 
\end{equation}
for each $i$.  Thus, (\ref{Fcon}) yields the tangent-space analogue (c$'$) of condition (c).

To summarize, the rigid, Euclidean geometry of $\bigl( T_{p_0} \M, g_{ab}(p_0) \bigr)$ suffices to prove the tangent-space analogues of our claims above.  More precisely, we have shown that 
\begin{enumerate}\renewcommand\theenumi{\alph{enumi}$'$}
\item the 1-parameter group of linear isometries $\Phi'(t)$ fixes the origin in $T_{p_0} \M$, \textit{whence}
\item every other orbit of $\Phi'(t)$ in $T_{p_0} \M$ is either a fixed point or a circle, and 
\item all of those circular orbits close for a common period $t = t_0$ as long as (\ref{Fcon}) holds.
\end{enumerate}
In addition, (\ref{Fcon}) holds identically if $\M$ is 2- or 3-dimensional.  Finally, note that the natural normalization of $F^b{}_a$ sets each $f_i = 1$, which is equivalent to 
\begin{equation}\label{Fnorm}
	F^3 = - F.
\end{equation}
This one normalization condition implies that \textit{all} orbits of $\Phi'(t)$ close at $t_0 = 2 \pi$.

\section{From Local to Global}

The results in the tangent space $T_{p_0} \M$ extend back to the whole manifold $\M$ essentially because the global isometries $\Phi(t)$ map geodesics of $(\M, g_{ab})$ to other geodesics.  Recall the exponential map 
\begin{equation}\label{expdef}
	\exp : T_{p_0} \M \to \M
	\qquad\text{via}\qquad
	\exp v := \gamma_v(1), 
\end{equation}
where $\gamma_v(s)$ denotes the unique geodesic of $(\M, g_{ab})$ starting from $\gamma_v(0) := p_0$, having initial tangent $\dot\gamma_v(0) := v \in T_{p_0} \M$, and affinely parameterized so that $\grad\!_{\dot\gamma_v(s)}\, \dot\gamma_v(s) = 0$ for all $s$.  Each global isometry $\Phi(t)$ maps this geodesic to another one, $\gamma_{\Phi'(t)(v)}(s)$, which of course has its initial tangent rotated in $T_{p_0} \M$.  In short, 
\begin{equation}\label{pComm}
	\Phi(t) \circ {\exp}(v) = {\exp} \circ \Phi'(t)(v)
\end{equation}
for all $t \in \Re$ and all $v \in T_{p_0} \M$.

If $\M$ is connected, compact, and without boundary, then there exists at least one geodesic between $p_0$ and any of its other points.  It therefore follows from (\ref{pComm}) that every orbit of $\Phi(t)$ in $\M$ must be the image under the exponential map of at least one orbit of $\Phi'(t)$ in $T_{p_0} \M$.  The exponential generally is not invertible, of course, so the orbit in $T_{p_0} \M$ generally is not unique.  But the orbits of $\Phi(t)$ cannot intersect one another, or themselves, non-trivially.  Thus, although the exponential mapping of orbits may be many-to-one, and may map circular orbits of $\Phi'(t)$ to fixed-point orbits of $\Phi(t)$, it nonetheless preserves orbits.  It follows in particular that each orbit of $\Phi(t)$ must either be a fixed point or a closed circle.  In other words, condition (b) for the orbits of $\Phi(t)$ follows from condition (b$'$) for the orbits of $\Phi'(t)$.

It is also clear that every circular orbit of $\Phi(t)$ must close with the same period as one of the orbits of $\Phi'(t)$ that is mapped to it by the exponential.  But all orbits of $\Phi'(t)$ have the same period as long as (\ref{Fcon}) holds, and condition (c) for the orbits of $\Phi(t)$ then follows from condition (c$'$) for the orbits of $\Phi'(t)$.  

Let us summarize the chain of arguments here.  Condition (a) on a Killing field on a Riemannian geometry $(\M, g_{ab})$ implies condition (a$'$) for the linear isometries $\Phi'(t)$ of the Euclidean geometry $\bigl( T_{p_0} \M, g_{ab}(p_0) \bigr)$.  This in turn implies conditions (b$'$) and (c$'$), as long as (\ref{Fcon}) holds.  Finally, these imply conditions (b) and (c) globally on $\M$ via the exponential map.  Note that the \textit{local} normalization condition (\ref{Fnorm}) at $p_0$ therefore dictates the \textit{global} scaling of the Killing field $\varphi^a$ such that all of its orbits have the desired period $t_0 = 2 \pi$.

\section{Application to a 2-Sphere}

Every vector field, and thus every Killing field, on a 2-sphere $S$ vanishes at at least one point, and (\ref{Fcon}) holds identically in 2 dimensions.  Every Killing field on a 2-sphere therefore satisfies conditions (a), (b), and (c).  But the intuitive picture of an axial Killing field on a 2-sphere is that it should vanish at exactly two poles, as in condition (A), not just at a single point.  We now show that this is necessarily the case.

Suppose that a 2-sphere geometry $(S, g_{ab})$ admits an axial Killing field $\varphi^a$, which we take to be normalized according to (\ref{Fnorm}).  Each point $p \ne p_0$ in $S$ lies on a unique orbit of $\varphi^a$, which divides the sphere into two regions.  Let $a(p)$ denote the area of the region bounded by that orbit and containing the fixed point $p_0$.  Equivalently \cite{AEPV:Mult}, we may set 
\begin{equation}\label{aGrad}
	\grad\!_b\, a := 2 \pi\, \epsilon_{bc}\, \varphi^c
	\qquad\text{with}\qquad
	a(p_0) := 0.
\end{equation}
The smooth function $a(p)$ takes both maximum and minimum values somewhere on the (compact) sphere, and each must occur at a fixed point where $\varphi^c$ vanishes.  The minimum value is zero, and occurs only at $p = p_0$.  The maximum value $A$ also occurs at a fixed point $p_1$, which cannot coincide with $p_0$ because any fixed point of $\varphi^a$ must be \textit{one-sided}.  That is, the local topology of $S^2$ is that of the 2-dimensional plane, so the fixed points of $\varphi^a$ must be isolated from one another, and each must be surrounded by a \textit{single} family of circular orbits.  (If two distinct families of circular orbits surrounded a given fixed point, then $S$ would have the local topology of a two-sided cone there.)  Similarly, every circular orbit of $\varphi^a$ must be \textit{two-sided} in the sense that there are exactly two distinct families of circular orbits surrounding it.  (There cannot be only one such family because $S$ has no boundary, and there cannot be more than two because this would violate the assumed manifold structure of $S$ at each point of such an orbit.)  The manifold of orbits $S / \Phi$ of $\varphi^a$ therefore has the topology of a closed interval $[0, A]$, with exactly two fixed points at the two ends and only circular orbits in the interior.  Thus, $\varphi^a$ must have exactly two isolated fixed points.

Finally, we show that the exponential map maps a closed disk with \textit{finite} radius $\sigma_1$ in $T_{p_0}$ onto \textit{all} of $S$, with the entire circular boundary of that disk mapped to the conjugate fixed point $p_1 \in S$.  This mapping is invertible in the interior of the disk, but obviously not on the boundary.  To see this, observe that 
\begin{equation}\label{gprop}
	\grad\!_{\dot\gamma_v(s)}\, \norm[\big]{\dot\gamma_v(s)} = 0
	\qquad\text{and}\qquad
	\grad\!_{\dot\gamma_v(s)}\, \bigl( \varphi \cdot \dot\gamma_v(s) \bigr) = 0
\end{equation}
because $\gamma_v(s)$ in (\ref{expdef}) is an affinely parameterized geodesic and $\varphi^a$ is a Killing field.  It follows that the tangent to $\gamma_v(s)$ has constant norm equal to $\norm{v}$ for all $s$ and, since $\varphi^a$ vanishes at $p_0$, is everywhere orthogonal to the circular orbits of $\varphi^a$.  The proper distance in $S$ along $\gamma_v(s)$ from $p_0$ to $p = \exp v$ is therefore equal to $\norm{v}$ in $T_{p_0} S$.  (This holds generally, not just on the 2-sphere.)  It also follows from (\ref{gprop}) that the proper geodetic distance from $p_0$ to $p_1$ is 
\begin{equation}\label{sigint}
	\sigma_1 
		:= \int_0^A \frac{\ed a}{\norm{\ed a}}
		= \int_0^A \frac{\ed a}{2 \pi \rho(a)}
		= \int_0^A \frac{\rho'(a)\, \ed a}{(\pi \rho^2)'(a)}, 
\end{equation}
where $\rho := \norm{\varphi}$ is the proper radius of the orbit of $\varphi^a$ through any given point, and primes denote derivatives with respect to $a$.  This integral is irregular because $\rho(a)$ vanishes at both endpoints.  But the integral converges nonetheless because the conditions to avoid conical singularities in $S$ at $p_0$ and $p_1$ are 
\begin{equation}\label{nocone}
	\pdby{a} \bigl( \pi \rho^2(a) \bigr) \Bigr|_{a = 0} = 1
	\qquad\text{and}\qquad
	\pdby{a} \bigl( \pi \rho^2(a) \bigr) \Bigr|_{a = A} = -1.
\end{equation}
Thus, although $\rho'(a)$ diverges at both endpoints in (\ref{sigint}), the indefinite integral of $\sigma(a)$ vanishes like $\rho(a)$ near $a = 0$.  The difference $\sigma_1 - \sigma(a)$ likewise vanishes like $\rho(a)$ near $a = A$.  There can be no other irregular points with $\rho(a) = 0$, so the integral (\ref{sigint}) is well-defined and $\sigma_1$ is finite.


\section{Conclusions}

We have seen that any Killing field $\varphi^a$ that vanishes at at least one point $p_0$ in a Riemannian manifold $(\M, g_{ab})$, and satisfies (\ref{Fcon}) there, necessarily has only fixed-point and closed, circular orbits throughout $\M$, the latter all having the same period $t_0$.  This offers a convenient way to characterize axial Killing fields in terms of local data at a fixed point.  Furthermore, scaling $\varphi^a$ globally such that (\ref{Fnorm}) holds locally at $p_0$ guarantees that the period of every circular orbit is $2 \pi$.

The condition (\ref{Fcon}) is not needed in 2 or 3 dimensions, and in fact one might imagine abandoning it altogether for manifolds of arbitrary dimension.  The main disadvantage of this would be that some orbits of $\Phi'(t)$ may no longer be closed.  Rather, if some of the $f_i$ from (\ref{fiVals}) were incommensurate, some orbits $\Phi'(t)$ would densely fill a torus in $T_{p_0} \M$.  The same would then apply to the orbits of $\Phi(t)$ by the same methods used above.  Although this would be rather less convenient, and less physical, such ``quasi-axial'' symmetries could potentially be interesting in certain applications.

One additional feature of these results is that they suggest a natural and convenient way to normalize an \textit{approximate} Killing field on a Riemannian geometry $(\M, g_{ab})$ that has no actual symmetries \cite{K:Conf, CW:Appr, LOPC:Eigen}.  Namely, suppose that a candidate approximate Killing field $u^a$ can be found up to constant scaling on $\M$ and vanishes at at least one point, which again is guaranteed if $\M \sim S^2$.  Then, one may define $F_{ab} := 2\, \grad\!_{[a}\, u_{b]}$, and use (\ref{Fnorm}) to normalize $u^a$ throughout $\M$ without having to compute the orbits of $u^a$ and ensure that they close, etc.  We explore this proposal further in a companion paper \cite{BW:Stab}.


\section*{Bibliography}

\begin{thebibliography}{MM}

\bibitem{BY:QLoc}
J.D.~Brown and J.W.~York,~Jr.  
Quasilocal energy and conserved charges derived from the gravitational action.
\textit{Phys.\ Rev.\ D}\ \textbf{47} (1993) 1407--1419.

\bibitem{AK:Rev}
A.~Ashtekar and B.~Krishnan.
Isolated and dynamical horizons and their applications.
\textit{Living\ Rev.\ Relativity} \textbf{7} (2004) 10.\\
\texttt{http://www.livingreviews.org/lrr-2004-10} (cited 18 Nov 2013)

\bibitem{ABL:Mech}
A.~Ahstekar, C.~Beetle and J.~Lewandowski.
Mechanics of rotating isolated horizons.
\textit{Phys.\ Rev.\ D} \textbf{64} (2001) 044016.

\bibitem{CLT:Unif}
X.~Chen, P.~Lu, and G.~Tian.
A note on uniformization of Riemann surfaces by Ricci flow.
\textit{Proc.\ Amer.\ Math.\ Soc.} \textbf{34} (2006), 3391--3393.

\bibitem{C:Sph}
B.~Chow.
The Ricci flow on the 2-sphere.
\textit{J.\ Differential Geom.} \textbf{33} (1991) 325-334.

\bibitem{AEPV:Mult}
A.~Ashtekar, J.~Engle, T.~Pawlowski and C.~Van Den Broeck.
Multipole moments of isolated horizons.
\textit{Class.\ Quantum Grav.} \textbf{21} (2004) 2549--2570.

\bibitem{K:Conf}
M.~Korzy\'nski.
Quasi-local angular momentum of non-symmetric isolated and dynamical horizons from the conformal decomposition of the metric.
\textit{Class.\ Quantum Grav.} \textbf{24} (2007) 5935--5943.

\bibitem{CW:Appr}
G.B.~Cook and B.F.~Whiting.
Approximate Killing vectors on $S^2$.
\textit{Phys.\ Rev.\ D} \textbf{76} (2007) 041501(R).

\bibitem{LOPC:Eigen}
G.~Lovelace, R.~Owen, H.P.~Pfeiffer and T.~Chu.
Binary-black-hole initial data with nearly extremal spins.
\textit{Phys.\ Rev.\ D} \textbf{78} (2008) 084017.

\bibitem{BW:Stab}
C.~Beetle and S.~Wilder.
Perturbative stability of the approximate Killing field eigenvalue problem.
Preprint \texttt{arXiv:??\ [gr-qc]}.

\end{thebibliography}
\end{document}